\documentclass[aps,prl,twocolumn,groupedaddress]{revtex4}
\usepackage{graphicx}

\newcommand{\xdir}{$\langle 1\overline{1} 0 \rangle$}
\newcommand{\ydir}{$\langle 110 \rangle$}

\newcommand{\xydir}{$\langle 100 \rangle$}

\newcommand{\Rxx}{$R_{xx}$}
\newcommand{\Ryy}{$R_{yy}$}
\newcommand{\Bp}{$B_{\perp}$}
\newcommand{\Bpar}{$B_{\parallel}$}

\begin{document}

\title{Metastable Resistance Anisotropy Orientation of Two-Dimensional Electrons 
in High Landau Levels}

\author{K.~B.~Cooper$^1$, J.~P. Eisenstein$^1$, 
L.~N. Pfeiffer$^2$, and K. W. West$^2$}

\affiliation{$^1$California Institute of Technology, Pasadena CA 91125 
\\
	 $^2$Bell Laboratories, Lucent Technologies, Murray Hill, NJ 
07974\\}


\begin{abstract} 
In half-filled high Landau levels, two-dimensional electron systems possess collective phases which exhibit a strongly anisotropic resistivity tensor.  A weak, but as yet unknown, rotational symmetry-breaking potential native to the host semiconductor structure is necessary to orient these phases in macroscopic samples.  Making use of the known external symmetry-breaking effect of an in-plane magnetic field, we find that the native potential can have two orthogonal local minima.  It is possible to initialize the system in the higher minimum and then observe its relaxation toward equilibrium.

\end{abstract}

\pacs{73.40.-c, 73.20.-r, 73.63.Hs}

\maketitle
There is now strong evidence for the existence of a broad new class of collective phases of two-dimensional electron systems (2DESs) which resemble charge density waves (CDWs) \cite{jpe1}.  These new phases, which have only been observed in very high quality samples and at very low temperatures, occur when several of the discrete Landau levels produced by a perpendicular magnetic field $B_{\perp}$ are occupied by electrons.  Their transport characteristics allow them to be readily distinguished from the fractional quantized Hall states at very high magnetic fields and the weakly disordered Fermi liquid at zero field. Most striking are those phases which develop around half-filling of the uppermost, or valence, Landau level (LL).  Corresponding to LL filling factors $\nu$ = 9/2, 11/2, 13/2, and so on, these phases are characterized by a huge anisotropy in the longitudinal resistance of the 2DES which develops rapidly on cooling below about 150 mK \cite{lilly1,du}.  This impressive signature has been widely interpreted as strong evidence for the unidirectional, or striped, CDW predicted via Hartree-Fock theory \cite{KFS,MC}.  In this theory, and its many extensions \cite{fogler}, at half-filling of high LLs the exchange-induced softening of the Coulomb interaction leads to phase separation of the 2DES into stripes in which the valence LL is alternately filled and empty.  Transport anisotropy is expected since it is almost certainly easier for charge to move parallel to the stripes than perpendicular to them. The period of the stripe pattern is estimated to be a few times the classical cyclotron radius $R_c$.

A key aspect of the anisotropic phases which lacks understanding is their consistent orientation with respect to the crystalline axes of the host GaAs material.  In virtually all samples this orientation is such that the high or ``hard" resistance direction is parallel to \xdir\ while the low or ``easy" direction is parallel to \ydir.  Numerous suggestions of weak native rotational symmetry-breaking potentials have been advanced, but so far none has been clearly shown to be relevant.  Early experiments demonstrated that a magnetic field component \Bpar\ in the plane of the 2DES provides an important tool for studying this potential.  Properly directed, an in-plane field can overcome the native symmetry-breaking potential and re-orient the resistive anisotropy \cite{pan,lilly2}.  A good understanding of this effect now exists \cite{tomas,phil} and, via comparison with the experiments, has shown that the strength of the native symmetry-breaking potential is only about 1 mK/electron \cite{cooper1}.  Here we present strong evidence that the native potential can in fact possess {\it two} distinct local minima, one favoring the hard resistance direction along \xdir\ and one favoring it along \ydir.  Which is the global minimum depends sensitively upon filling factor and in-plane magnetic field.  We find it possible to prepare the system in a metastable state of the ``wrong" orientation and observe its slow relaxation toward the equilibrium state determined via a field-cooling technique. 

Figure 1a shows the longitudinal resistance vs. perpendicular magnetic field $B_\perp$ of a high mobility 2DES at $T$ = 100 mK.  The 2DES resides in a 30 nm GaAs quantum well and has a density and low temperature mobility of $N_s=3\times 10^{11} ~\rm{cm^{-2}}$ and $\mu=3\times 10^7 ~\rm{cm^2/Vs}$, respectively.  Resistance measurements are made, via conventional lock-in techniques, using ohmic contacts positioned at the corners and side midpoints of the 5 mm square sample.  For the solid trace the average current flow is along \xdir\ and the resulting resistance is denoted \Rxx, while for the dotted trace the current flow is along \ydir\ and the resistance is \Ryy. (These conventions will be followed throughout this paper.)  The data shown in Fig. 1a cover LL filling factors $\nu = hN_s/eB_{\perp}$ between $\nu =4$ and $\nu =8$, corresponding to the Fermi level lying in the spin-split third (N=2) and fourth (N=3) orbital LLs.  In spite of the relatively high temperature at which these data were obtained, there is much evidence of collective phenomena in Fig. 1a.  For example, 
near $B_{\perp} \approx 2.6$ and 2.9T both \Rxx\ and \Ryy\ vanish.  Although these features (indicated by upward arrows in the figure) occur at fractional filling factors ($\nu \approx 4.72$ and 4.28, respectively) they in fact reflect {\it integer} QHE states \cite{lilly1,du}.  Distinct from the conventional integer QHE states at $\nu =4$ and 5, these new states are widely believed to be pinned triangular ``bubble" CDWs closely related to the unidirectional stripe phases at half filling. 

\begin{figure}
\centering
\includegraphics[width=3.25 in,bb=143 194 416 541]{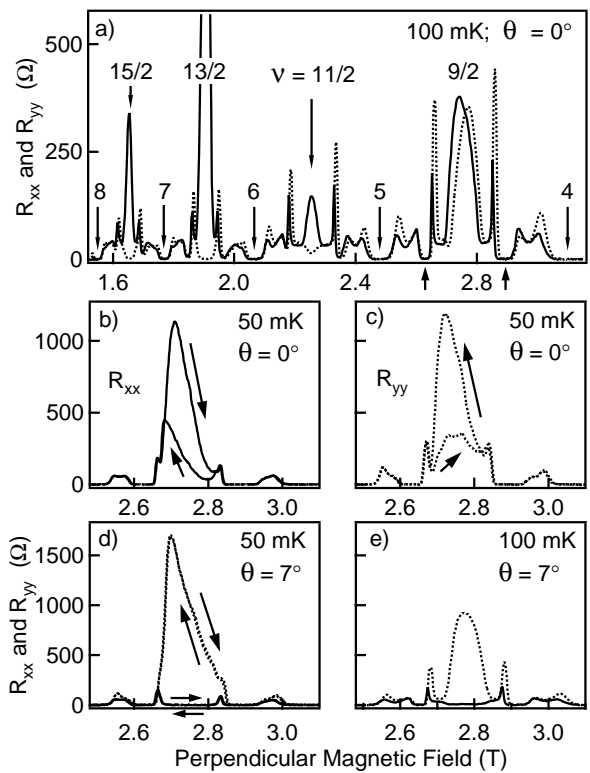}
\caption{\label{fig:fig1} a) Longitudinal resistances \Rxx (solid) and \Ryy\ (dotted) at $T$ = 100 mK.  b) and c) Sweep direction dependences of \Rxx\ and \Ryy\ at $T$ = 50 mK around $\nu = 9/2$. d) and e) Sample tilted by $\theta = 7^{\circ}$.  \Rxx\ and \Ryy\ show a strong and non-hysteretic anisotropy at $\nu = 9/2$ at $T$ = 50 and 100 mK. }
\end{figure}

More importantly, around filling factors $\nu = 11/2$ in the N=2 LL and $\nu = 13/2$ and 15/2 in the N=3 LL, the longitudinal resistance exhibits the strong anisotropy associated with the putative stripe phases, with \xdir, as usual, being the hard direction.  Interestingly, however, near $\nu = 9/2$, where anisotropy is usually strongest, the resistances \Rxx\ and \Ryy\ are almost identical.  As we will show, this observation does not imply a lack of local stripe order, but instead indicates that here the native symmetry-breaking potential has two nearly degenerate minima along the \xdir\ and \ydir\ directions.

Figures 1b and 1c display the resistances \Rxx\ and \Ryy, respectively, in the region around $\nu = 9/2$ at the lower temperature of 50 mK.  Each panel contains two traces; one taken with the magnetic field sweeping up at $dB/dt = 0.01$ T/min and one with the field sweeping down at the same rate.  Unlike the situation at 100 mK, both \Rxx\ and \Ryy\ exhibit a strong dependence upon the direction of the field sweep at 50 mK.  This dependence is confined to the region around half-filling; in the flanks of the LL, including the re-entrant integer QHE states, no sweep direction dependence is seen.  These data make it clear that hard and easy transport directions cannot be assigned at $\nu = 9/2$: On sweeping the field up the hard direction appears to be along \xdir; on sweeping down it seems to be \ydir.  A similar effect is seen in the region around $\nu =11/2$, even though at 100 mK \xdir\ is clearly the hard axis.

Figures 1d shows that the application of a small in-plane magnetic field completely removes the sweep direction dependence of the resistance around $\nu = 9/2$.  To obtain these data the sample was tilted by $\theta = 7^{\circ}$ around the \xdir\ direction; at $\nu = 9/2$ this produces an in-plane field of $B_{||} = 0.34$ T directed along \ydir.  In this situation the hard axis of the resistive anisotropy is unambiguously \ydir, in agreement with earlier tilted field measurements \cite{pan,lilly2}. Fig. 1e demonstrates that the small applied in-plane field is sufficient to bring out a strong anisotropy at $\nu = 9/2$ even at 100 mK. 

Similar observations have been reported by Zhu, {\it et al.} \cite{zhu}. In their experiments the 2DES density was swept at fixed magnetic field.  A strong dependence of the longitudinal resistance on the direction of the density sweep was observed at $\nu = 9/2$ if the density was near $N_s = 2.9 \times 10^{11} \rm cm^{-2}$.   For densities well below this value the hard direction at $\nu = 9/2$ was found to be \xdir, while for higher densities it was \ydir. Zhu, {\it et al.} did not propose an explanation for this density-driven re-orientation of the resistive anisotropy and suggested that their results could not rule out either a unidirectional or a bidirectional native symmetry-breaking potential.  

Our experiments, which employ a sample with a fixed density close the critical one identified by Zhu, {\it et al.} \cite{zhu}, strongly point to the existence of a bidirectional potential. The sweep direction dependence of the resistivity shown in Figs. 1b and 1c proves that over a range of magnetic fields about $\nu = 9/2$ there are two orthogonal orientations simultaneously possible for the resistive anisotropy \cite{no45}.  These two orientations likely correspond to two distinct minima of the symmetry-breaking potential.  Apparently the relative depth of these two minima is a sensitive function of filling factor.  For magnetic fields slightly below $\nu = 9/2$, the data in Figs. 1b and 1c imply the potential minimum for stripes with hard axis along \xdir\ is lower in energy.  Conversely, for fields slightly above $\nu = 9/2$, the potential favors the hard axis along \ydir. If the time required for stripe domains to relax from one minimum to the other is sufficiently long, then this picture would naturally explain the sweep direction dependence of the resistances.  The elimination of this hysteresis by an in-plane magnetic field is a consequence of the in-plane field's own symmetry-breaking effect favoring one of the two minima of the native potential.

\begin{figure}
\centering
\includegraphics[width=3.25 in,bb=137 170 419 521]{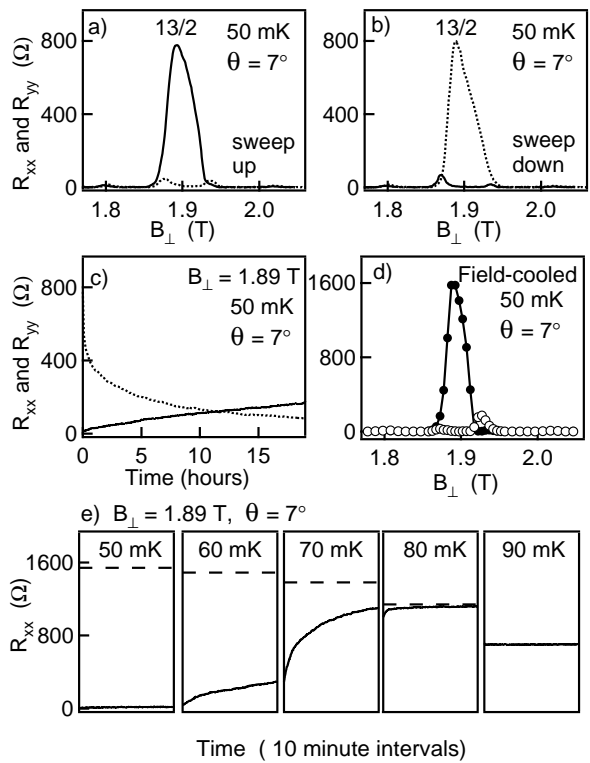}
\caption{\label{fig:fig1} a) and b) Resistances \Rxx\ (solid) and \Ryy\ (dotted) near $\nu = 13/2$ at $T$ = 50 mK showing extreme sweep direction dependence. Sample tilted by $\theta = 7^\circ$. c) Time evolution of \Rxx\ and \Ryy\ following field sweep {\it down} to \Bp = 1.89 T at 50 mK. d) \Rxx\ (solid dots) and \Ryy\ (open dots) after cooling from 150 to 50 mK at fixed magnetic field. e) Initial time dependences of \Rxx\ following field sweep down to \Bp =1.89 T at 50 mK. Dashed lines are field-cooled values of \Rxx.}
\end{figure}

In order to make this model more convincing we now turn to the behavior of the resistance near $\nu = 13/2$.  Here we have directly observed the slow relaxation of metastable anisotropy orientations toward an equilibrium configuration.  Moreover, a field-cooling technique allows for unambiguous identification of the equibrium configuration, avoiding the complications of hysteresis and slow relaxation times.  

The data in Fig. 1a show that at $T$ = 100 mK and $B_{||} = 0$ a large resistive anisotropy is present at $\nu = 13/2$, with the hard direction along \xdir.  On cooling to 50 mK we find that a slight magnetic field sweep direction dependence does develop, but the hard transport direction remains \xdir\ regardless of the sweep direction.  Within our model of a bidirectional symmetry-breaking potential, this implies that while the relative depth of the two local minima varies with magnetic field around $\nu = 13/2$, which minimum is lowest in energy does not change.  The small hysteresis observed presumably arises from a small non-equilibrium population of stripe domains with the ``wrong" orientation.  If this is the case, the application of an appropriate in-plane magnetic field ought to bring the two minima closer together and exacerbate the sweep direction dependence.  As Figs. 2a and 2b show, this is precisely what we observe. With the sample once again tilted to $\theta = 7^{\circ}$, producing $B_{||} = 0.23$ T along \ydir\ at $\nu = 13/2$, the observed sweep dependence of the resistive anisotropy is extreme.  A large anisotropy of very similar aspect is seen in both Fig. 2a and 2b, only the hard and easy directions are interchanged.  As with $\nu =9/2$, sweeping the magnetic field up favors the hard direction along \xdir, while sweeping it down favors \ydir. 

If the sweep direction dependence of the resistive anisotropy is due to striped domains getting trapped in the ``wrong" direction as the magnetic field is swept into the region of half-filling, owing to a filling factor dependence of the relative depth of the two potential minima, then it is reasonable to expect to observe temporal relaxation of the resistance, provided that the energy barrier between the two wells is not too large.  Figure 2c shows that just such relaxation is observed.  In the figure \Rxx\ and \Ryy\ at $T$ = 50 mK are plotted vs. time, over a 19 hour interval. These data were obtained by first sweeping the magnetic field {\it down} to $B_{\perp}$ = 1.89 T, stopping, and then recording the subsequent time evolution.  This stopping field was chosen because it is in the center of the anisotropic region about $\nu =13/2$.  Immediately after stopping at 1.89 T, $R_{yy} \approx 800 ~\Omega$, while \Rxx\ is only about  $\approx 7 ~\Omega$.  Subsequently \Ryy\ falls quickly at first and then slows down. Meanwhile, \Rxx\ slowly rises.  After about 11 hours the two resistances cross and \xdir\ becomes the hard direction.  As time evolves further, the gap between \Rxx\ and \Ryy\ grows steadily.  These data clearly suggest that the initial anisotropy orientation present when the field sweep was stopped (hard axis along \ydir) was a metastable one and that the equilibrium orientation has the hard axis along \xdir.  

To confirm this scenario we have employed a field-cooling technique to determine the equilibrium configuration of the anisotropy.  In this technique the magnetic field is swept into the $\nu =13/2$ vicinity at $T$ = 150 mK.  At this high temperature no anisotropy is evident in the resistances and no hysteresis is observed.  Once the field sweep is stopped, the sample is allowed to cool to $T$ = 50 mK and both \Rxx\ and \Ryy\ are then recorded.  Figure 2d shows the results of this procedure.  The solid dots represent the $T$ = 50 mK values of \Rxx; the open dots \Ryy.  We emphasize that no time dependence of these values is observed after cooling to 50 mK and no dependence upon the direction of the 150 mK field sweep is found. It is clear from Fig. 2d that for most of the magnetic field range around $\nu =13/2$, the equilibrium hard direction is \xdir.  Only in a very narrow range on the high field side of the anisotropic region does \ydir\ appear to assume that role. 

That the field-cooled resistances shown in Fig. 2d are the equilibrium ones is demonstrated in Fig. 2e.  Each panel displays both the time dependence of \Rxx\ over a 10 minute interval following a downward field sweep to \Bp\ = 1.89 T at $T$ = 50 mK, and the field-cooled value (dashed horizontal lines) of the same resistance obtained via the field-cooling technique described above.  The data show that the \Rxx\ values observed after the 50 mK field sweeps do indeed relax toward the equilibrium values found after cooling from 150 mK at fixed field. 
Not surprisingly, the relaxation time is strongly temperature dependent: It is many hours at 50 mK, several minutes at 70 mK, and almost immediate at 90 mK.
Assuming the relaxation rate is determined by thermal activation over the barrier between the two potential minima, a rough estimate of the barrier height is $U_b \sim 1$ K.

\begin{figure}
\centering
\includegraphics[width=3.25 in,bb=0 10 255 186]{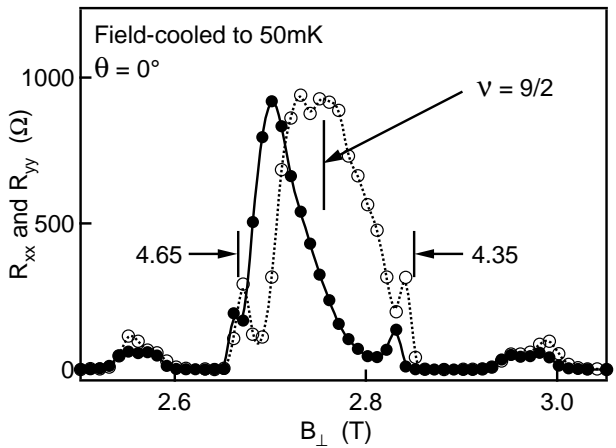}
\caption{\label{fig:fig1} \Rxx\ (solid dots) and \Ryy\ (open dots) in vicinity of $\nu = 9/2$ after field-cooling to $T$ = 50 mK.  Sample is untilted.}
\end{figure}

Applying the field-cooling technique to the resistances around $\nu = 9/2$ in the N = 2 LL yields the data in Fig. 3.  These 50 mK data were obtained with $B_{||} = 0$ and therefore only the native symmetry-breaking potential is present.  These equilibrium results further support our model of a bidirectional native symmetry-breaking potential.  Only a modest level of resistive anisotropy is present in the window $4.35 < \nu < 4.65$ centered on half-filling. For $4.35 < \nu < 4.57$, \Ryy\ exceeds \Rxx, while for $4.57 < \nu < 4.65$ \Rxx\ exceeds \Ryy.  These results suggest that the two orthogonal minima of the native potential are nearly degenerate throughout the 9/2 region, a conclusion consistent with the lack of anisotropy at $T$ = 100 mK evident in Fig. 1a.  Moreover, the hard resistance direction being along \ydir\ for magnetic fields above $\nu =4.57$ and along \xdir\ for fields below this point, makes sense of the hysteresis data shown in Figs. 1b and 1c.  Sweeping into the 9/2 region from high fields would initially generate striped domains with hard axes along \ydir, while approaching from low fields would initially produce domains with hard axes along \xdir.  Given the demonstrated long relaxation times of metastable anisotropy orientations at low temperatures, hysteresis is obviously expected. 

The existence of two orthogonal local minima in the native anisotropy potential, one favoring the hard axis along \xdir\ and one favoring \ydir\, implies that in addition to a term in the native potential proportional to cos$(2\phi)$ (with $\phi$ being the angle between the hard transport direction and \xdir), there is also a term proportional to cos$(4\phi)$.  The sign of the cos$(2\phi)$ term determines whether \xdir\ or \ydir\ is the preferred hard axis direction. Prior experiments \cite{zhu} suggest that the sign of this term changes at a critical density.  Our results are consistent with this but reveal additional dependences on LL index and the precise filling within each spin-resolved LL.  The physical origin of the cos$(2\phi)$ term remains unknown \cite{cooper1}.  Unlike the cos$(2\phi)$ term, a cos$(4\phi)$ term is not prohibited by the crystal symmetry of bulk GaAs.  Our results demonstrate that this term leads to dramatic metastability phenomena when the cos$(2\phi)$ term is small.  The latter occurs when the density is near the critical one \cite{zhu} or when an appropriate in-plane magnetic field is applied. One interesting recent suggestion \cite{piezo} is that the piezoelectricity of GaAs produces a cos$(4\phi)$ term with minima favoring \xdir\ and \ydir.  The barrier separating these degenerate minima is roughly 1 mK/electron.  Comparing this with our crude estimate $U_b \sim 1$ K for the height of the barrier which inhibits relaxation of mis-oriented stripe domains leads us to speculate that each domain contains about $10^3$ electrons, corresponding to lateral dimensions of order $1\mu$m. 

In conclusion, we have presented evidence that the native symmetry-breaking potential responsible for orienting the anisotropic phases near half-filling of high Landau levels has two orthogonal local minima, one favoring the hard resistance direction along \xdir\ and one favoring \ydir.  We have shown that it is possible to witness the dynamic relaxation of the anisotropy from metastable states to an equilibrium orientation separately determined via a field-cooling technique.

We thank E.I. Rashba for a helpful correspondence.  This work was supported by the DOE under Grant No. DE-FG03-99ER45766 and the NSF under Grants No. DMR-0070890 and DMR-0242946.


\begin{thebibliography}{99}
\bibitem{no45} We stress that the resistance anisotropy does not appear to ``rotate" smoothly between \xdir\ and \ydir.  For example, it is never seen along \xydir.
\end{thebibliography}
\end{document}